\documentclass[pss]{wiley2sp} % provides pss two-column style
\usepackage{amsmath,latexsym}
\usepackage{color}

\newcommand{\half}{\mbox{$\textstyle \frac{1}{2}$}}
\newcommand{\ket}[1]{\left | \, #1 \right \rangle}
\newcommand{\kets}[1]{| \, #1 \rangle}
\newcommand{\bra}[1]{\left \langle #1 \, \right |}

\newcommand{\brackets}[3]{\langle\, #1\,|#2|\,#3\,\rangle}

\newcommand{\av}[1]{\langle #1\rangle}
\newcommand{\outprod}[2]{\ket{#1}\bra{#2}}

\newcommand{\eqr}[1]{Eq.~\ref{#1}}

\begin{document}

% Title of the article
\title{Capturing the re-entrant behaviour of one-dimensional Bose-Hubbard model}

% Abbreviated title for the page headers
\titlerunning{Capturing the re-entrant behaviour}

% Authors
\author{%
  M.~Pino\textsuperscript{\textsf{\bfseries 1}},
  J.~Prior\textsuperscript{\textsf{\bfseries 2}},
  S.R.~Clark\textsuperscript{\Ast,\textsf{\bfseries 3,4}}}

% Abbreviated list of authors for the page headers
\authorrunning{M.~Pino et al.}

%E-mail-address of corresponding author
\mail{e-mail \textsf{s.clark@physics.ox.ac.uk}, Phone: +44-1865-272388, Fax: +44-1865-272400}

% author's affiliations/addresses
\institute{%
  \textsuperscript{1}\,Departamento de F\'{\i}sica - CIOyN, Universidad de Murcia, Murcia 30071, Spain\\
  \textsuperscript{2}\,Departamento de F\'{\i}sica Aplicada, Universidad Polit\'ecnica de Cartagena, Cartagena 30202, Spain\\
  \textsuperscript{3}\,Centre for Quantum Technologies, National University of Singapore, 2 Science Drive 3, Singapore 117542\\
  \textsuperscript{4}\,Clarendon Laboratory, University of Oxford, Parks Road, Oxford OX1 3PU, U.K.}

\received{XXXX, revised XXXX, accepted XXXX} % do not change, will be filled in by the publisher
\published{XXXX} % do not change, will be filled in by the publisher

% Please select about four verbal keywords for your manuscript.
\keywords{Re-entrance, Bose-Hubbard model, Mott Insulator, Superfluid}

\abstract{
\abstcol{
The Bose Hubbard model (BHM) is an archetypal quantum lattice system exhibiting a quantum phase transition between its superfluid (SF) and Mott-insulator (MI) phase. Unlike in higher dimensions the phase diagram of the BHM in one dimension possesses regions in which increasing the hopping amplitude can result in a transition from MI to SF and then back to a MI. This type of re-entrance is well known in classical systems like liquid crystals yet its origin in quantum systems is still not well understood. Moreover, this unusual re-entrant character of the BHM is not easily captured in approximate analytical or numerical calculations. }{Here we study in detail the predictions of three different and widely used approximations; a multi-site mean-field decoupling, a finite-sized cluster calculation, and a real-space renormalization group (RG) approach. It is found that mean-field calculations do not reproduce re-entrance while finite-sized clusters display a precursor to re-entrance. Here we show for the first time that RG does capture the re-entrant feature and constitutes one of the simplest approximation able to do so. The differing abilities of these approaches reveals the importance of describing short-ranged correlations relevant to the kinetic energy of a MI in a particle-number symmetric way.}}

\maketitle   % please do not remove

\section{Introduction}
Re-entrance is a novel phenomenon in which a succession of transitions between two phases A and B, such as A-B-A-B, can occur by monotonically increasing just one parameter. Such a series of transitions implies a non-linearity which allows a single parameter to produce opposite effects depending on its initial value and is often a reflection of the interacting character of the system. In the context of classical thermal phase transitions the natural parameter to vary is temperature. It is then expected that the low temperature phase A will be ordered, while increasing the temperature will drive the system to a disordered phase B. However, the appearance of re-entrance means that increasing the temperature can in fact unexpectedly stabilize the ordered phase A again. This sequence of phase transitions has been observed in liquid crystals between the A = Smectic (ordered) phase and B = Nematic (disorded) phase with increasing temperature~\cite{Cladis75,Chandrasekhar92}, and similar behaviour also occurs in frustrated classical spin systems~\cite{Saslow86}.

The appearance of re-entrance in the zero-temperature phase diagram of a quantum lattice system is much less well understood. In this work, we study the Bose-Hubbard model (BHM) which is a minimal many-body Hamiltonian where the competition between the kinetic and repulsive interaction effects gives rise to a quantum phase transition~\cite{Sachdev01}. For weak interactions the bosons are completely delocalized leading to a superfluid (SF) phase, while for sufficiently strong interactions, and a commensurate density, the bosons become localized and enter the Mott insulator (MI) phase. The BHM has long been used to describe Josephson junction arrays in which several superconducting island are coupled together~\cite{Bruder05}. More recently it has received considerable attention and experimental prominence due to its near perfect realization with cold atoms trapped in optical lattices~\cite{Jaksch98,Bloch08,Greiner02,Stoferle04,Spielman07,Gemelke09}. 

The qualitative structure of the BHM phase diagram, such as the existence of MI lobes depicted in Fig.~\ref{fig1}(a), was worked out some time ago by Fisher {\em et al}.~\cite{Fisher89}. They showed that the tip of the lobes undergo a transition belonging to the universality class of the {\em XY} model in $d+1$ dimensions, implying that in one dimensions (1D) it is a Kosterlitz-Thouless (KT)~\cite{Kosterlitz73} transition. Only much later was it discovered that the MI lobes in 1D actually displays re-entrance at low filling factors. Specifically at some constant chemical potential, near the tip of the lobe, a re-entrant sequence of zero-temperature quantum phase transitions occur, with A = MI and B = SF, as the coherent hopping amplitude is increased from zero. This is again surprising since it demonstrates that increasing the hopping amplitude, which in isolation favours the itinerancy of the bosons, can instead favour their localization under certain circumstances. This delicate behaviour is not easily captured in commonly used approximations, a fact connected to the known difficulty of describing the tips KT transition. Instead the existence of re-entrance was firmly established using sophisticated numerical tools such as density matrix renormalization group (DMRG)~\cite{Kuhner98,Kuhner00}, high-order strong-coupling perturbation (SCP) expansions~\cite{Freericks96,Elstner99b}, and quantum Monte Carlo (QMC) simulations~\cite{Batrouni90,Batrouni92,Kashurnikov96a}. However, an understanding of the underlying mechanism leading to re-entrance has yet to be achieved. Our aim here is to glean insight into the key physical properties required for its emergence in the BHM in 1D by analyzing the capabilities of relatively simple and intuitive approximation methods to capture it. 

The structure of this paper is as follows. In Sec.~\ref{sec:BHM}, we discuss the essential properties of the MI and SF phases and employ several qualitative arguments to understand the structure of the BHM phase diagram. This is followed by a discussion of the exact form of phase diagram for the BHM in 1D, displaying re-entrance, as determined by earlier DMRG, SCP expansions, and QMC calculations. Then in Sec.~\ref{sec:CR} three different widely used approximation methods are applied to the BHM. This includes a multi-site mean-field decoupling~\cite{Sheshadri93,Rokhsar91,Krauth92,Sachdev01,Buonsante05,Pisarski11}, a finite-sized cluster calculation~\cite{Park04,Zhang10,Elesin94,Buonsante07} and a renormalization group (RG) approach~\cite{Singh92}. Finally, in Sec.~\ref{sec:Conclusions} several conclusions are summarized from the capabilities of these methods at describing re-entrance.

\section{Bose-Hubbard model} \label{sec:BHM}
The BHM describes a bosonic system in a regular lattice where particles can tunnel to nearest neighbors and interactions occurs between bosons at the same lattice site. The Hamiltonian in the grand canonical ensemble is (taking $\hbar = 1$)
\begin{eqnarray}
\hat{H} &=& -t\sum_j\left(\hat{b}^{\dagger}_j\hat{b}_{j+1} + \textrm{h.c.}\right) -\mu \sum_j\hat{n}_j \nonumber \\
&& + \frac{U}{2}\sum_j\hat{n}_j(\hat{n}_j-1), \label{eq:bhm_ham}
\end{eqnarray}
where $\hat{b}_j,\hat{b}^{\dagger}_j$ are the bosonic creation and annihilation operators at site $j$ and $\hat{n}_j = \hat{b}_j^\dagger\hat{b}_j$ is the corresponding number operator. Here $t>0$ is the hopping amplitude of bosons between neighbouring sites, $\mu$ is the chemical potential and $U>0$ is the repulsive on-site interaction strength. In the non-interacting limit, $U=0$, the Hamiltonian \eqr{eq:bhm_ham} is diagonal in the basis of single-particle plane waves with a tight-binding dispersion relation $\epsilon(k) = -2t\cos(ka)$, where $a$ is the lattice spacing. In this case, the ground-state is characterized by all the particles condensing into the $k=0$ state. The number of particles in the ground-state diverges whenever $\mu<-2t$. In the following we fix our energy scale by $U$ and consider the phase diagram in the $(t,\mu)$ plane where for simplicity we label the dimensionless ratios as $t/U \rightarrow t$ and $\mu/U \rightarrow \mu$.

\begin{figure*}[t!]
\begin{center}
\includegraphics[width=13cm]{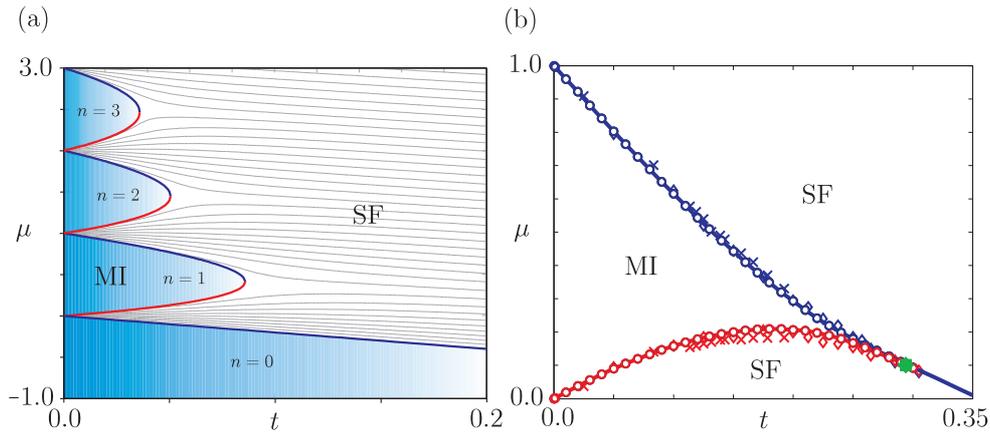}
\caption{(a) The mean-field phase $(t,\mu)$ diagram of the BHM in one dimension~\cite{Sheshadri93,Rokhsar91,Krauth92}. The shaded regions are the MI lobes with integer fillings $n$. These parabolic shaped lobes are surrounded by the SF phase. Some density contours in the SF phase are shown to illustrate their tendency to have a negative slope. Note also that the MI lobes are particle-hole asymmetric. The hole boundary lower side of the lobe while the particle boundary is upper side. (b) Focusing on the $n=1$ MI lobe the DMRG results from Ref.~\cite{Kuhner00} are shown as $\circ$, along with two different sets of QMC data from Ref. \cite{Batrouni92} as $\times$ and Ref.~\cite{Kashurnikov96a} as $\Diamond$. The solid lines delineate the region found to have unit density using a 12th order SCP expansions~\cite{Elstner99b} and the large solid point marks the KT tip found from DMRG~\cite{Kuhner00}.} \label{fig1}
\end{center}
\end{figure*}

In the $t=0$ limit, the energy eigenstates possess a well defined occupation at each lattice site. The ground state is uniformly filled with $n$ bosons in each site giving a commensurate density so long as the chemical potential fulfills~\cite{Sachdev01}
\begin{eqnarray}
n-1<\mu<n. \label{eq.occupation}
\end{eqnarray}
In the case $\mu>0$, the ground state at zero hopping is filled with at least one boson per site. Excitations above this state can be constructed by adding (particle excitations) or removing (hole excitations) a boson to the ground state at any site. The energy of these lowest particle and hole excitation with respect to the ground state energy is $\Delta_{p}\left(\mu\right)=-\mu+n$ and $\Delta_{h}\left(\mu\right)=\mu-n+1$. For a non-integer $\mu$, we see from Eq.~(\ref{eq.occupation}) that $\Delta_{p}\left(\mu\right)>0$ and $\Delta_{h}\left(\mu\right)>0$. Thus, a defining characteristic of the MI phase is the gap induced by the interaction for both particles and hole excitations~\cite{Wen07}.

Keeping $\mu>0$ constant, the particle and hole gap decreases when hopping is increased from the $t=0$ limit. Depending on $\mu$, there is a hopping value at which one of the particle or hole gaps becomes zero. At this point, the system undergoes a phase transition to the gapless SF phase. Commonly the system goes from a commensurate to an incommensurate density when generically crossing the MI-SF phase transition. We refer to the phase boundary in which the particle gap closes as the particle boundary and as the hole boundary when the hole gap closes. Crossing the phase boundaries from the MI to the SF region produces an increase (particle boundary) or decrease (hole boundary) in the density. 

In 1D quantum fluctuations destroy long-range order in the SF phase. Instead, this phase exhibits quasi-long range order with algebraic decaying correlations, such as $\av{\hat{b}^\dagger_0\hat{b}_x} \sim x^{-K/2}$ with $K = \pi/\sqrt{\rho_s\kappa t}$, where $\rho_s$ is the superfluid density~\cite{Giamarchi04} and $\kappa=\partial\rho/\partial\mu$ is the compressibility~\cite{Giamarchi04}. This lack of genuine long-range order shows that the SF phase in 1D is not a Bose-Einstein condensate. Nevertheless, it is the non-vanishing $\rho_s$ that defines this phase as a genuine SF~\cite{Annett04}. Within the MI phase correlations decay exponentially fast with a correlation length $\xi$, such as $\av{\hat{b}^\dagger_0\hat{b}_x} \sim \exp(-x/\xi)$. These correlations indicate that the bosons in a MI state are localized over a typical region of length $\xi$ contrasting to the SF regime where particles can move over the whole system. 

In the SF region, the constant-density region are curves in the $(\mu,t)$ plane due to the need for positive compressibility, $\kappa>0$. Thus, the particle and hole boundaries surrounding each of the MI regions must collapse in a constant-density curve in the SF phase resulting in the well known MI lobe structure~\cite{Sachdev01,Fisher89}. It also implies that the MI to SF transition at the tip of each of the MI lobes then occurs at constant density coinciding with the merging of the particle and hole boundaries. The scaling theory developed by Fisher {\em et al.}~\cite{Fisher89} showed that generic transitions are in the universality class of the Gaussian model, while the transition occurring at the tips belongs to the universality class of the $XY$-model in $d+1$ dimension. Thus, a KT critical point appears at the tips of each MI lobe in 1D problem so both particle and hole gap close exponentially slowly as $\Delta(t) =A\exp(-B/\sqrt{t_{\rm KT} - t})$, with $A,B$ real constants~\cite{Kosterlitz73}.

A qualitative description of the phase diagram for the 1D BHM can be obtained using a single-site MF decoupling approximation~\cite{Sheshadri93,Rokhsar91,Krauth92,Sachdev01}. This results in the phase diagram shown in Fig.~\ref{fig1}(a). The light-blue area indicates the commensurately filled MI lobes while the white denote SF regions surrounding them. Also shown are some density contours in the SF phase. Since in the non-interacting limit, equivalent to $t\rightarrow \infty$, the particle number diverges for any $\mu$, this implies that the density contours in the SF must have a negative slope whenever the hopping is large enough. As can be seen in Fig.~\ref{fig1}(a) this feature is reproduced by the MF approximation. This kinetic effect occurring at moderate values of hopping can be viewed as a precursor to re-entrance~\cite{Kuhner00}, as we shall see shortly. However, this simple single-site MF approximation, which is essentially equivalent to the infinite dimensional limit, fails to predict re-entrant MI lobes. 

\section{Capturing re-entrance} \label{sec:CR}
Re-entrance was not predicted for the 1D BHM until more sophisticated and accurate approximations were applied. In particular it was first revealed by using 12th order SCP expansions~\cite{Elstner99b}. After this discovery further calculations with DMRG~\cite{Kuhner00} and QMC methods~\cite{Batrouni92,Kashurnikov96a} convincingly established the phenomenon. Their original results are replotted in Fig.~\ref{fig1}(b) showing the most accurate, and essentially exact, result for the 1D BHM phase diagram. A clear consensus from these calculations is that the MI lobe in 1D and for small filling factor $n\sim 1$, has a triangular pointed shape which is quite different from the rounded shape revealed in the MF approximation. Moreover the tip of the lobe, also highlighted in Fig.~\ref{fig1}(b) from the DMRG calculation~\cite{Kuhner00}, occurs at a much larger value of $t$ than MF predicts owing to the KT nature of its transition in 1D. Roughly speaking re-entrance occurs due to a combination of this elongation of the tip with the kinetic energy driven tendency for all density contours to eventually have a negative slope. With this view it is therefore also natural that re-entrance only occurs in 1D because in higher dimension the transition at the tip of the MI lobe is not a KT critical point and instead occurs at values of $t$ increasingly close to the MF prediction. Ultimately these two effects cause the hole boundary to become concave before it intersects with the particle boundary. This then makes it possible to traverse, by increasing $t$ alone, a series of phases transitions MI-SF-MI-SF at constant $\mu$ in the region just above the tip. That re-entrance occurs in this region, where the interplay between hopping and interactions are most significant, illustrates how strong correlations are key to its appearance. We note also that increasing the filling $n$ causes bosonic enhancement of the kinetic energy and consequently the MI lobe shrinks as $\approx 1/n$. Since the tip then occurs at much smaller values of $t$ re-entrance disappears for $n>1$. For this reason we focus entirely on $n=1$ filling in this work.

The sophisticated methods employed to determine this structure however, obscure the underlying physical intuition behind re-entrance owing to their complexity. Our main aim here is to capture re-entrance within the simplest possible approximation in order to gain further insight into the origin of this remarkable phenomenon. We now examine in turn some common approaches.

\subsection{Multi-site mean-field decoupling} \label{sec:MSMF}
The MF decoupling approach is often a highly successful means of determining the qualitative structure of the phase diagram for quantum lattice systems.  For the BHM the most effective strategy is to decouple the hopping term via the factorization~\cite{Sheshadri93,Sachdev01}
\begin{equation}
\hat{b}_i^\dagger\hat{b}_j \approx \hat{b}_i^\dagger \av{\hat{b}_j} + \av{\hat{b}_i^\dagger}\hat{b}_j - \av{\hat{b}_j^\dagger}\av{\hat{b}_i},
\end{equation}
where $\av{\cdot}$ denotes the ground state average. Applying this factorization leaves the interaction term untouched but reduces the BHM to a sum of identical single site Hamiltonians  
\begin{eqnarray}
\hat{H}_{1} &=&-2t(\Phi \hat{b}^{\dagger} + \Phi^*\hat{b}) +
2t|\Phi|^2 + \frac{U}{2}\hat{n}(\hat{n}-1)-\mu\hat{n}. \nonumber
\end{eqnarray}
It is known that the exact solution of the SF in 1D does not posses genuine long-range order implying that $\av{\hat{b}_j}=0$. However, this form of mean-field approximation tacitly assumes that $\av{\hat{b}_j}$ can be non-zero and so breaks the global $U(1)$ symmetry of the BHM while preserving its translational symmetry. In fact this complex scalar provides with a homogeneous order parameter $\Phi =\av{\hat{b}_j}$ for all sites $j$ accounting for the ``mean-field'' effect of hopping between neighbouring sites. This is formally identical to performing a variational calculation using a translationally invariant Gutzwiller ansatz~\cite{Krauth92,Rokhsar91} for the ground state of the form
\begin{equation}
\ket{\Psi_1} = \prod_{j}\ket{\psi_1}_j, 
\end{equation}
over all sites $j$. Here $\ket{\psi_1}$ is the ground state of the single-site Hamiltonian $\hat{H}_1$ which can be determined straightforwardly by numerical diagonalization once an appropriate occupancy cut-off is introduced~\cite{Sheshadri93,Sachdev01}. By iterating this procedure a self-consistent solution is found where the $\Phi$ appearing in $\hat{H}_1$ is equal to the expectation value $\bra{\psi_1}\hat{b}_j\ket{\psi_1}$ of its ground state and is equivalent to minimizing the  ground state energy $\epsilon_{1}(\Phi)$ with respect to $\Phi$. The MI phase is identified by a symmetry preserving  ground state $\ket{\psi_1}$ with $\Phi=0$ and so its description is limited to integer filled Fock states only, whilst the SF phase is described by a symmetry breaking superposition of on-site Fock states with $\Phi\neq 0$. Due to the product nature of the ansatz $\ket{\Psi_1}$ no quantum correlations between sites are described by this approximation. 

\begin{figure}[t!]
\begin{center}
\includegraphics[width=7cm]{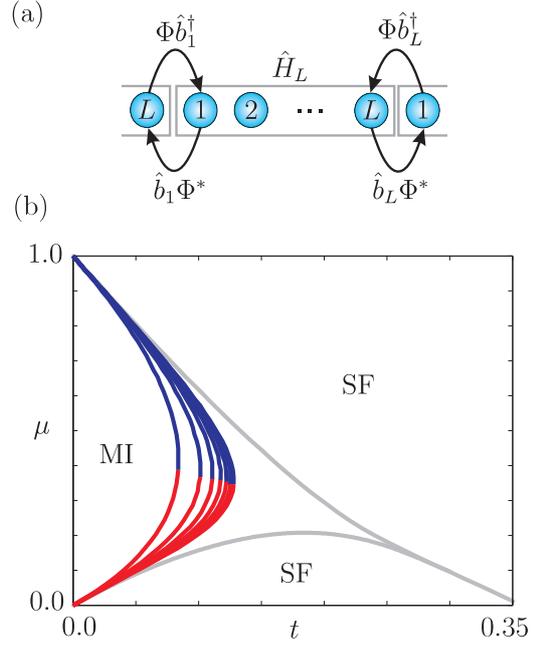}
\caption{(a) A block of $L$ sites is represented together with its nearest neighbours from the two contiguous blocks. The hopping between the two sites at the end of the central block with the sites outside of it are approximated using a homogeneous SF order parameter $\Phi =\av{\hat{b}_1} = \av{\hat{b}_L}$ which is determined self-consistently~\cite{Buonsante05,Pisarski11}. In doing so, the Hamiltonian of the central block is decoupled from its neighbours sites rendering the system numerically solvable. (b) Several MI lobes are shown which were computed using mean-field decoupling on the hopping between adjacent 1D blocks of $L$ sites. The size of the block vary from the conventional $L=1$ up to $L=8$ site clusters. The solid gray lines are the boundaries from a 12th order SCP expansion which are included as a reference.} \label{fig2}
\end{center}
\end{figure}

The rounded MI lobes predicted by the conventional single-site MF was shown earlier in Fig.~\ref{fig1}(a). We now examine the results produced by a multi-site generalization of this approach~\cite{Buonsante05,Pisarski11}. Specifically, we apply the same decoupling approximation to the hopping terms connecting the boundaries of each $L$-site blocks as shown schematically in Fig.~\ref{fig2}(a). A virtue of this approach is that the decoupled systems now have a 1D chain geometry and so the approximation departs from the effective infinite dimensional geometry of the single-site scheme. The MF decoupling is equivalent to assuming a product ansatz
\begin{equation}
\ket{\Psi_L} = \prod_{\mathcal{C}}\ket{\psi_L}_\mathcal{C}, 
\end{equation}
over all $L$-site blocks $\mathcal{C}$, with where $\ket{\psi_L}$ is the ground state of the cluster Hamiltonian
\begin{eqnarray}
\hat{H}_{L}&=&-t(\Phi \hat{b}_1^{\dagger} + \textrm{h.c.})  -t(\Phi \hat{b}_L^{\dagger} + \textrm{h.c.})  + 4t|\Phi|^2-\mu\sum_{j=1}^L\hat{n}_j \nonumber \\
           & & -t\sum_{j=2}^{L-1}\left(\hat{b}^{\dagger}_j\hat{b}_{j+1} + \textrm{h.c.}\right) + \frac{U}{2}\sum_{j=1}^L
\hat{n}_j(\hat{n}_j-1). \nonumber
\end{eqnarray}
Here the boundary MF is $\Phi =\av{\hat{b}_1} = \av{\hat{b}_L}$ enforcing a homogeneous order parameter whose value is determined self-consistently as before. The MI and SF phase are again signalled by $\Phi = 0$ and $\Phi\neq 0$, respectively. By partitioning the system in to clusters of $L>1$ sites our description loses its full translational symmetry, but in turn can describe MI and SF phases by more complex states with quantum correlations, such as particle-hole fluctuations, present within each block. 

As seen in Fig.~\ref{fig2}(b) we find that the extension to a multi-site MF approach has only a marginal influence on both the shape and size of the predicted MI lobes. Going to large chains does extend the lobe to fractionally larger $t$, though the qualitative form of the MI lobes remain rounded similar to the single-site version. In principle as $L \rightarrow \infty$ this approach will converge to an exact diagonalization of the BHM. However, the results for numerically accessible chains sizes up to $L=8$ show this convergence is extremely slow and no re-entrant behaviour is visible. This suggests that the symmetry breaking induced by the boundary MF decoupling of the hopping is dominant and washes out vital physics underpinning re-entrance.

\subsection{Small system exact diagonalization}\label{sec:SSED}
\begin{figure}
\begin{center}
\includegraphics[width=7cm]{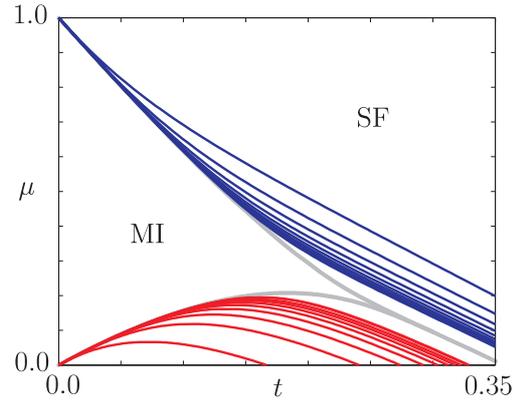}
\caption{The sequence of particle and hole boundaries computed from small finite-sized clusters with periodic boundary conditions~\cite{Park04,Buonsante07}. System sizes go from $N=2$ to $N=11$ sites as the outermost to innermost lines, respectively. The solid gray lines are the boundaries for a 12th order SCP expansion which are included as a reference.} \label{fig3}
\end{center}
\end{figure}

Despite the fact that no finite-sized system possesses a genuine phase transition the exact diagonalization of small systems can nonetheless provide important information about physics in the thermodynamic limit~\cite{Park04,Zhang10,Elesin94,Buonsante07}. Utilizing symmetries and the sparseness of the BHM Hamiltonian, with an on-site occupancy cut-off of 4 bosons and periodic boundary conditions, we applied the Lanczos algorithm to numerically diagonalize system sizes from $N=2$ up to $N=11$ sites~\cite{Zhang10}. To examine the phase diagram we employed a quasi-grand-canonical approach. By working in the canonical ensemble for an $N$ site system we computed, as a function of $t$, the ground state energy $\epsilon_{N+1}(t)$, $\epsilon_{N}(t)$ and $\epsilon_{N-1}(t)$ for the system with exactly $N+1$, $N$ and $N-1$ bosons, respectively. We then extract an effective grand-canonical ensemble phase diagram by computing the particle and hole boundaries in the $(t,\mu)$ plane as the degeneracy curves between these particle number sectors given by the solutions to the equations
\begin{eqnarray}
\epsilon_{N+1}(t) - \epsilon_N(t) + \mu &=& 0, \quad \textrm{(particle)}, \nonumber \\
\epsilon_{N}(t) - \epsilon_{N-1}(t) + \mu &=& 0. \quad \textrm{(hole)} \nonumber
\end{eqnarray}
In Fig.~\ref{fig3} the resulting particle and hole boundaries for the MI lobe for increasing system sizes $N$ are shown. An indication that these curves do not produce genuine phase boundaries is that the particle and hole boundaries do not intersect for any finite $N$ and so predict a MI lobe of infinite size. However, a key feature, clearly visible for all system sizes examined, is that the hole boundary already shows concavity at moderate values of $t$ characteristic of re-entrance. Moreover there is a visible tendency of the finite-size system boundaries to slowly converge to the expected MI lobe in the thermodynamic limit as predicted by the 12th order SCP expansion. This feature is the basis of finite-size scaling calculations~\cite{Park04,Buonsante07}.

Both the finite-sized cluster and the multi-site MF approximations are based on an ansatz with a repeated cluster of variable size which breaks the underlying translational symmetry of the BHM Hamiltonian in Eq.~(\ref{eq:bhm_ham}). Yet the phase boundaries predicted for each of these methods are rather different. We see that the finite-sized cluster calculation produces systematic overestimations of the MI lobe for any cluster size while the multi-site MF approximation produces a systematic underestimation. The reason for this is that unlike MF decoupling the finite-sized cluster approach does not break the $U(1)$ global symmetry of the BHM and is unable to mimic the quasi-long range order characteristic of a SF thereby favouring the MI phase. In contrast the multi-site MF approximation can break this symmetry favouring the SF phase. That a precursor to re-entrance is visible in finite-sized clusters demonstrates that a symmetry preserving description of kinetic energy is crucial to its emergence in an approximation. This conclusion is further enhanced by results obtained with more sophisticated approaches such as the variational cluster perturbation method~\cite{Koller06}. There the system is partitioned into finite-size clusters, like that used in the multi-site MF approximation, however hopping between the clusters is instead treated perturbatively. Crucially this allows the $U(1)$ symmetry to be preserved in both the SF and MI phases. This phase diagram produced by this method then resemble the exact-diagonalization results here and also similarly predict re-entrance for any cluster size~\cite{Koller06}.

\subsection{Real-space renormalization group}\label{sec:RSRG}
The exponential growth in the Hilbert space with $N$ means that exact diagonalization cannot be feasibly pushed much beyond the sizes considered here already. For this reason we examine the predictions of a real-space RG approach for the BHM following the scheme originally introduced by Singh and Rokhsar~\cite{Singh92} some time ago. In that work they focused on the canonical ensemble at unit-filling to analyse the KT transition point at the tip. Here for the first time we reformulate their scheme within the quasi-grand-canonical ensemble approach, just applied to the finite-sized clusters, to obtain an RG prediction of the complete phase diagram of the BHM. The RG approximation combines two crucial features of the MF and finite-sized clusters considered already. First, like finite-sized clusters it preserves the $U(1)$ global symmetry of the BHM. Second, like the MF decoupling RG scheme curtails the exponential growth of the system's Hilbert space enabling the thermodynamic limit to be reached.  

The real-space RG scheme applied is based on partitioning the 1D system into non-overlapping two site blocks. In contrast to the other approximations the RG scheme~\cite{Singh92} simplifies the BHM by truncating the on-site Fock basis to the occupation states $\ket{0}, \ket{1}$ and $\ket{2}$ thereby excluding triple or higher occupancies. In this basis the kinetic contribution $\hat{K}$ to the block Hamiltonian $\hat{H}_b$, describing hopping between adjacent sites within a block, is
\begin{eqnarray}
\hat{K} &=& -t_h \big( \outprod{0\,1}{1\,0} + \textrm{h.c.}\big) \nonumber \\
&& -t_o \big(\outprod{1\,1}{2\,0} + \outprod{1\,1}{0\,2} +  \textrm{h.c.}\big) \nonumber\\
&& -t_p \big( \outprod{2\,1}{1\,2} +  \textrm{h.c.}\big) \nonumber
\end{eqnarray}
where $t_h$ and $t_p$, specify the hopping amplitude for a hole or additional particle, respectively, while $t_o$ is the amplitude for the creation or annihilation of a particle-hole pair. When constructing
 $\hat{K}$ directly from the hopping terms in the BHM in Eq.~(\ref{eq:bhm_ham}) we have that $t_h = t_o/\sqrt{2} = t_p/2 = t$ reflecting the fact that the BHM is not particle-hole symmetric. The block Hamiltonian $\hat{H}_b$ is then completed by adding the interaction and chemical potential terms, which in the truncated basis are diagonal on-site operators with diagonal matrix elements $(0,0,U)$ and $(0,\mu,2\mu)$, respectively~\cite{Singh92}.

The RG scheme begins by diagonalizing $\hat{H}_b$ and identifying three eigenstates $\ket{0'}, \ket{1'}$ and $\ket{2'}$ which represent the lowest energy state with one, two and three bosons per block, respectively. The Fock state truncation enables these states to be readily determined as
\begin{eqnarray}
\ket{0'} &=& \frac{1}{\sqrt{2}}\big(\ket{0\,1} + \ket{1\,0}\big), \nonumber \\
\ket{1'} &=& \cos(\theta)\ket{1\,1} + \frac{1}{\sqrt{2}}\sin(\theta)\big(\ket{0\,2} + \ket{2\,0}\big), \nonumber \\
\ket{2'} &=& \frac{1}{\sqrt{2}}\big(\ket{1\,2} + \ket{2\,1}\big), \nonumber
\end{eqnarray}
with corresponding energies
\begin{eqnarray}
E_{0'} &=& -\mu - t_h, \nonumber \\
E_{1'} &=& -2\mu - \sqrt{2} \, t_o\sin(2\theta) + U\sin^2(\theta), \nonumber \\
E_{2'} &=& -3\mu - t_p + U, \nonumber
\end{eqnarray}
where $\theta = \half \tan^{-1}(2\sqrt{2}\, t_o/U)$. Using this appropriately chosen reduced basis of states for a single block we can now interpret the block as a new super-site and formulate a renormalized block Hamiltonian $\hat{H}'_b$ spanning two such adjacent super-sites. In this way the RG scheme dramatically limits the degrees of freedom retained. The Hamiltonian $\hat{H}'_b$ has exactly the same type of terms as $\hat{H}_b$, once the replacements $\ket{0} \rightarrow \ket{0'}$, etc. have been made, but possesses different matrix elements $t'_h, t'_o, t'_p, U'$ and $\mu'$. The new on-site matrix elements are given by 
\begin{eqnarray}
\mu' &=& E_{1'} - E_{0'} = \mu - t_h[1-2\sin(2\theta)] + U\sin^2(\theta), \nonumber \\
U' &=& E_{2'} + E_{0'} -2E_{1'} = U\sec(2\theta) - t_p - t_h. \nonumber 
\end{eqnarray}
The new hopping amplitudes are then determined by computing directly in the two block Hilbert space the matrix elements
\begin{eqnarray}
t'_h &=& -\brackets{0'\,1'}{\hat{K}}{1'\,0'}, \nonumber \\
t'_o &=& -\brackets{1'\,1'}{\hat{K}}{2'\,0'} =  -\brackets{1'\,1'}{\hat{K}}{0'\,2'}, \nonumber \\
t'_p &=& -\brackets{2'\,1'}{\hat{K}}{1'\,2'}, \nonumber
\end{eqnarray}
where $\hat{K}$ is the hopping term connecting the two blocks~\cite{Singh92}. Repeating this process for $k$ RG iterations yields matrix elements $t^{(k)}_h, t^{(k)}_o, t^{(k)}_p, U^{(k)}, \mu^{(k)}$ as well as energies $E_0^{(k)},E_1^{(k)},E_2^{(k)}$ and their corresponding states $\kets{0^{(k)}}, \kets{1^{(k)}}, \kets{2^{(k)}}$. 

\begin{figure*}[t!]
\begin{center}
\includegraphics[width=13cm]{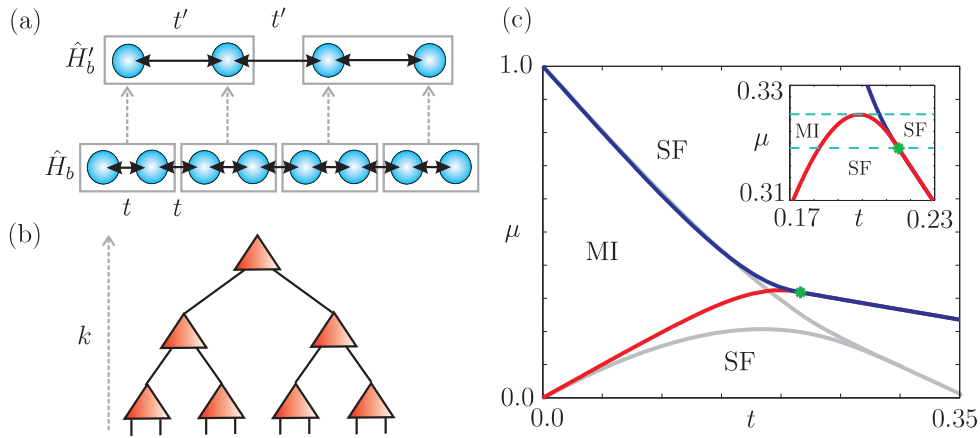}
\caption[]{(a) The real-space RG scheme adopted here~\cite{Singh92}. The Hamiltonian of a two site block is diagonalized and the lowest one, two and three boson eigenstates are kept. This forms an isometry mapping $3\otimes 3 \rightarrow 3$ which renormalizes a block of two sites into a single super-site coupled to it neighbours with a different hopping, i.e. $t \rightarrow t'$. (b) This RG scheme is equivalent to variationally minimizing a TTN ansatz. Here each isometry is represented as a 3 leg tensor (triangle) composed as a hierachical network. (c) The MI lobe predicted by the real-space RG scheme is displayed, along with the green point marking the predicted KT critical point \cite{Singh92}. The inset is a zoom of phase boundaries around the tip highlighting the observed re-entrance. The solid gray lines are again the boundaries from a 12th order SCP expansion included as a reference.} \label{fig4}
\end{center}
\end{figure*}

Typically after $k \approx 100$ RG iterations the scheme drives the system to particle-hole symmetry with $t^{(k)}_h = t^{(k)}_o = t^{(k)}_p = t^{(k)}$ and converges to one of two fixed points. The MI is signified by $(t/U)^{(k)} = 0$ whereupon $\kets{1^{(k)}} = \kets{1^{(k-1)} \, 1^{(k-1)}}$ illustrating that the RG has removed the short-range correlations and converged to a unit-filled Fock state. The SF is signified by the ratio $(t/U)^{(k)} = 0.5165$ possessing non-zero particle-hole fluctuations. The repeated renormalization of two-site blocks is equivalent to variationally approximating the ground state of the BHM via a hierarchically structured tensor tree network (TTN), as schematically shown in Fig.~\ref{fig4}(b). As with the multi-site MF decoupling and finite-sized clusters this TTN ansatz breaks the translationally symmetry of the BHM.

The RG ansatz possesses crucial similarities to a finite-sized cluster approach. If after $k$ RG steps the single open leg at the top of the TTN was capped by the final block state $\kets{1^{(k)}}$ the resulting network is a highly efficient representation of a many-body state of an $N=2^k$ site chain filled with exactly $N$ bosons. It therefore preserves the $U(1)$ global symmetry. Similarly capping the TTN with $\kets{0^{(k)}}$ or $\kets{2^{(k)}}$ yields a representation of a state with $N-1$ and $N+1$ bosons, respectively. Thus, while the RG procedure can easily reach the thermodynamic limit it can only describe very limited $\pm 1$ particle-number fluctuations due to the repeated truncation applied. This means that the RG approach might capture correlations of a MI reasonably, but can only offer a very poor description of the SF phase. 

In an analogous way to the finite-size cluster approximation, the RG prediction for the phase diagram was determining by numerically solving for the degeneracy curves $E_1^{(k)} - E_0^{(k)} = 0$ and $E_2^{(k)} - E_1^{(k)} = 0$ in the $(t,\mu)$ plane between $\kets{1^{(k)}}$ and the states $\kets{0^{(k)}}$ and $\kets{2^{(k)}}$, respectively. The resulting hole and particle boundaries are reported in Fig.~\ref{fig4}(c). The point where they meet forms the lobe tip whose location is a unstable fixed point that can be found analytically to be at $(t/U)^{(k)} = 0.215$~\cite{Singh92}. The predicted MI lobe has a triangular shape and from inset of Fig.~\ref{fig4}(c), zoomed in near the tip, the appearance of re-entrance is evident. Compared to the SCP expansion result the particle boundary is in reasonable agreement, however the KT point is under-estimated and the hole boundary differs noticeably away from the $t \rightarrow 0$ limit. Nonetheless the RG approximation has captured the re-entrant feature of the MI lobes in 1D and its success strongly points to a $U(1)$ symmetric description of particle-hole fluctuations of a MI state as being essential. This conclusion is consistent with recent investigations linking re-entrance in the BHM in 1D to entanglement using a more sophisticated infinite matrix product ansatz~\cite{Pino12}. 

\section{Conclusions}\label{sec:Conclusions}
Re-entrance is often difficult to capture in approximations of the 1D BHM. Nevertheless, we have shown that this real-space RG scheme does in fact reproduce this phenomenon despite being much simpler than the DMRG, high order SCP expansions or QMC. The failures of a multi-site MF decoupling and finite-sized cluster approach reveal clues as to why the RG approximation has this ability. Specifically, it retains some, albeit limited, capacity to describe in a particle-number symmetric way relevant fluctuations of the MI while truncating sufficiently to enable the thermodynamic limit to be approached. This, and the precursors to re-entrance seen in finite-sized clusters, highlights the importance of short range correlations in the MI state near the tip. Furthermore, the truncation of the on-site occupancy in the RG scheme shows that particle-hole pair correlations in particular, arising from kinetic energy, play a significant role. This can be contrasted with the related issue of the location of the KT critical point where long-range correlations are instead crucial. An interesting open question is how an extended real-space RG scheme, where the initial on-site Fock basis and subsequent RG steps truncate less, improves upon the simplest approach used here while retaining a physically intuitive form. Our work here suggests that a moderate improvement in the description of particle-hole fluctuations could substantially improve the predicted MI lobe shape and KT point. 

\begin{acknowledgement}
MP acknowledges support from DGI Grant No. FIS2009-13483. JP was supported by Ministerio de Ciencia e Innovaci\'on Project No. FIS2009-13483-C02-02 and the Fundaci\'on S\'eneca Project No. 11920/PI/09-j. SRC thanks the National Research Foundation and the Ministry of Education of Singapore for support.  
\end{acknowledgement}

\end{document}